\documentclass[doublecol,figures]{epl2}
\usepackage{graphicx}
\usepackage{ulem}
\usepackage{dcolumn}
\usepackage{amsmath}
\usepackage{amssymb}
\usepackage{color, colordvi}
\usepackage{float}
\usepackage{placeins}
\usepackage{epstopdf}
\usepackage{subfigure}

\def\beq{\begin{equation}}
\def\eeq{\end{equation}}

\title{Anomalous Fisher-like zeros for the canonical partition function of noninteracting fermions }
\author{Rajat K. Bhaduri\inst{1} \and Allison MacDonald\inst{1} \and Wytse van Dijk\inst{1,2}} 
\institute{\inst{1} Department of Physics and Astronomy,  
McMaster University, Hamilton, Canada L8S 4M1 \\
\inst{2} Physics Department, Redeemer University College, Ancaster, Ontario, Canada L9K 1J4}

\abstract{
Noninteracting fermions, placed in a system with a continuous density 
of states, may have zeros in the $N$-fermion canonical partition function 
on the positive real $\beta$ axis (or very close to it), even for a small
number of particles. This 
results in a singular free energy, and instability  
in other thermal properties of the system. In the context of trapped
fermions in a harmonic oscillator, these zeros are shown to be
unphysical.  Our results are also applicable to mean-field canonical calculations for fermions.   By contrast, similar bosonic calculations with continuous density of states yield sensible results.  
}


\begin{document}

\maketitle

Long back, Lee and Yang~\cite{yang52,lee52} pointed out that the onset of
a phase transition could be deduced by studying the zeros of the
grand partition function on the complex fugacity plane. As the number of
particles goes to infinity in the thermodynamic limit, the complex zeros
tend to pinch the real fugacity axis, signalling a phase transition. 
Later, Fisher~\cite{fisher65} found a similar behaviour 
of the canonical partition function $Z_N(\beta)$ on the complex 
$\beta$ (inverse temperature) plane near a phase transition. Given a 
single-particle partition function $Z_1(\beta)$, 
$Z_N(\beta)$ may be calculated exactly for noninteracting bosons or 
fermions using recursion relations~\cite{borrmann93}.  The single-particle partition function may arise from a one-body trapping potential, or may be the result of a calculation in a many-body problem. 
In this context, M\"{u}lken {\it et al.}~\cite{mulken01} 
have studied the pattern of Fisher
zeros for trapped noninteracting bosons in a harmonic oscillator. As
the number of bosons was increased, the real positive $\beta$ axis
tended to be pinched at $T=T_c$, the Bose-Einstein condensation temperature for spatial dimensions $d>1$.
The behaviour of the heat capacity as a 
function of $T$ also showed a sharp peak for large $N$ at $T=T_c$. 
As expected, this was found both for the exact
single-particle density of states of a harmonic oscillator (HO),
and for  
the corresponding (asymptotic) continuous density of states. 
M\"{u}lken {\it et al.}~\cite{mulken01} examined the distribution of
  zeros on the complex $\beta$ plane to classify the order of the
  phase transition in finite systems. Similar analyses were 
done with the Fisher zeros in interacting 
statistical models by Janke {\it et al.}~\cite{janke01,janke02}. 

Noninteracting fermions trapped in a HO have not been studied in this context, presumably because no irregularity or phase transition is expected.   
However, a classical version of this formalism has been applied in a model for the thermodynamic properties of nuclear multifragmentation in heavy-ion collisions~\cite{das05}.  In this paper, we are especially interested in the effects of
the Fermi statistics on the analytical properties of the canonical 
partition function. In particular, we compare results obtained from
the exact single-particle partition function from  the 
discrete HO energy spectrum, with those obtained
using a continuous density of states approximation. The latter is
widely used in the grand canonical formalism.
With the exact HO density of states, 
the fermionic system shows no evidence of any phase transition in any 
dimension, as expected. With continuous density of states, however, we 
obtain some very surprising results, to be described below. 
We conclude that for fermions trapped in a HO,  
or, for that matter, confined in a box, these zeros are spurious.  
We should mention here that in quantum field theory, 
lattice calculations of free Wilson fermions also manifest a real zero 
on the analogous inverse coupling axis if periodic boundary conditions
are used~\cite{kenna00}. This also gives rise to anomalous 
properties of the system~\cite{baille87}. 
Our paper may stimulate the search for spurious zeros of the canonical partition function in other areas of physics.  They may appear when a continuous density of states is used that relaxes the restrictions in long-range correlations, or the use of improper boundary conditions.  The long-range correlations in our case arose from the Pauli principle, but in other cases may conceivably arise from long-range interactions. 

The $N$-particle canonical partition function is generated from a one-body partition function $Z_1(\beta)$ by using a recursion relation~\cite{borrmann93}
\beq
Z_N(\beta)=\frac{1}{N}\sum_{k=1}^{N} \sigma^{k+1}Z_1(k\beta) 
Z_{N-k}(\beta)~,
\label{party}
\eeq
where $\sigma=1$ for bosons, and $(-1)$ for fermions, and $Z_0(\beta)=1$.   
The particles are (otherwise) taken to be noninteracting and identical, but correlations coming from quantum statistics are included. 
In this paper, for analytical simplicity, the particles are taken to be noninteracting, and in a $d$-dimensional isotropic 
harmonic oscillator.  The corresponding one-body partition function is 
given by~\cite{brack03}  
\beq
Z_1(\beta)  =\left[\sum_{n=0}^{\infty}
  e^{\textstyle -(n+1/2)\hbar\omega\beta}\right]^d  
   = \left[\frac{1}{2\sinh(\hbar\omega\beta/2)}\right]^d~,
\label{sinh}
\eeq
where we have not included the spin degeneracy for fermions. This
$Z_1(\beta)$, when substituted in eq.~(\ref{party}), enables us to 
calculate $Z_N(\beta)$ exactly for noninteracting bosons or fermions 
occupying the discrete HO spectrum of energies. 
The single-particle density of states, $g(\epsilon)$ may be obtained
from a Laplace inversion of $Z_1(\beta)$, and may be written as a sum 
of a smoothly varying part $\widetilde{g}(\epsilon)$, and an infinite sum
of oscillatory terms~\cite{brack03}. For example, for $d=2$,  
$g^{(2)}(\epsilon)=\Sigma_{n=0}^{\infty}(n+1)
\delta(\epsilon-(n+1)\hbar\omega)$, and may be written exactly as 
\beq
g^{(2)}(\epsilon)=\frac{\epsilon}{(\hbar\omega)^2}\left[1+2\sum_{k=1}^{\infty}
\cos\left(2\pi k\frac{\epsilon}{\hbar\omega}\right)\right].
\label{osc}
\eeq 
To see how this comes about, note that~\cite{philips57} 
\begin{equation}\label{eq:a}
Z_1(\beta) 
 = \dfrac{1}{4} \mathrm{cosech}^2\left(\dfrac{\hbar\omega\beta}{2}\right) 
= \dfrac{1}{4}\sum_{n=-\infty}^\infty \dfrac{1}{\left(\dfrac{\hbar\omega\beta}{2} + i n\pi\right)^2}~,
\end{equation}
with no restriction on $\beta$. On taking the Laplace inverse with
respect to $\epsilon$, the single
pole at $\beta=0$ for $n=0$ yields the first term on the right-hand
side of eq.~(\ref{osc}) , and the two poles for every integer $n>0$ at 
$\beta=\pm 2in\pi/\hbar\omega$ give rise to the oscillatory terms. 
The continuous part of the density of states above is 
$\epsilon/(\hbar\omega)^2$, which we denote by 
$\widetilde{g}^{(2)}(\epsilon)$. In thermodynamic calculations, it is the 
continuous density of states $\widetilde{g}^{(d)}$ that is used 
widely\footnote{This smooth density of states could also be derived by a semiclassical
expansion of $Z_1(\beta)$ in powers of $\hbar$~\cite{abramowitz65}. Note that  
\begin{equation}
\mathrm{cosech} (z) = \dfrac{1}{z} - \dfrac{z}{6} + \dfrac{7}{360} z^3 - \cdots , \ \ |z|<\pi.  \nonumber
\end{equation}
Putting $z= \hbar\omega\beta/2$, we get
\begin{equation}\label{eq:d}
Z_1(\beta) = \dfrac{1}{4}
\mathrm{cosech}^2\left(\dfrac{\hbar\omega\beta}{2}\right) =
\dfrac{1}{(\hbar\omega\beta)^2} - \dfrac{1}{12} +
\dfrac{(\hbar\omega\beta)^2}{240} - \cdots, \nonumber 
\end{equation}
with the restriction that $|\hbar\omega\beta|<2\pi $ 
for the asymptotic series to be valid. 
On Laplace inverting with respect to $\beta$, the smooth 
$\widetilde{g}(\epsilon)$ is again reproduced, followed by
a delta function and its derivatives at the origin.}.
The smooth one-body (Thomas-Fermi) partition function in $d$ dimensions is 
\beq
\widetilde{Z}_1(\beta)=\frac{1}{(\hbar\omega\beta)^d}~, 
\label{cont}
\eeq
whose Laplace inverse yields the continuous density of states 
\beq
\widetilde{g}^{(d)}(\epsilon)=\dfrac{\epsilon^{d-1}}{(d-1)!(\hbar\omega)^d}~. 
\label{smoothy}
\eeq

\begin{figure}[htbp]
  \centering
   \resizebox{3.0in}{!}{\includegraphics[angle=-90]{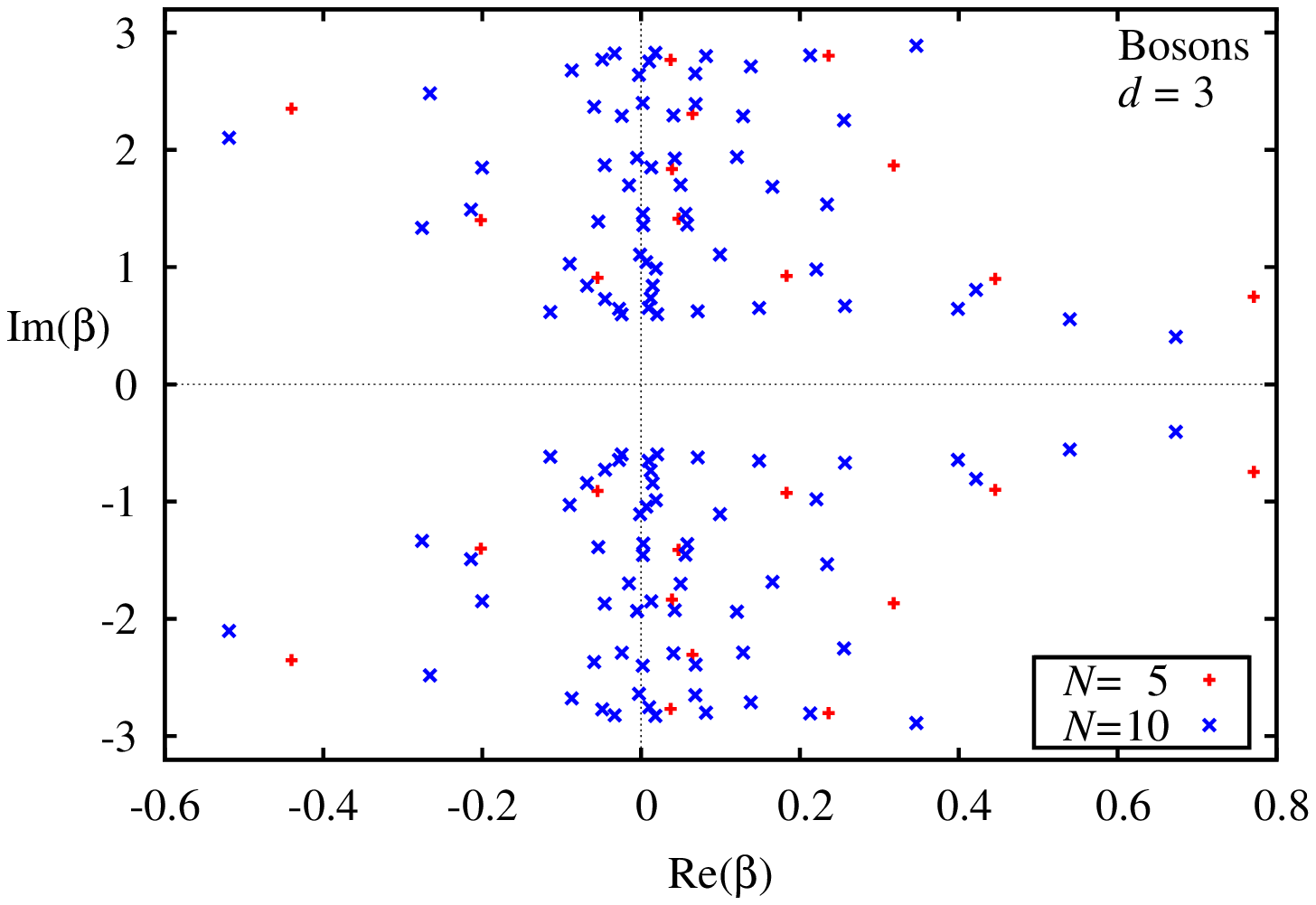}}
 \caption{(Color online) The zeros of the exact $Z_N(\beta)$, for 
$N=5$ and $N=10$, obtained from $Z_1(\beta)$
  of eq.~(\ref{sinh}) as a function of $\beta=1/kT$ for noninteracting
  bosons in a three-dimensional harmonic oscillator well. 
}
\label{fig:0}
\end{figure}

The zeros of $Z_N(\beta) $
on the complex $\beta$ plane may be calculated for bosons as well as
fermions. Note that the analytical expression for  $Z_N(\beta)$ is  
determined by the choice of $Z_1(\beta)$ given by eq.~(\ref{sinh}), or 
its continuous counterpart $\widetilde{Z_1}(\beta)$ defined above. Schmidt and 
Schnack~\cite{schmidt99} have expressed $Z_N(\beta)$ in terms of
polynomials in $\exp (-\hbar\omega\beta)$, which make the calculation
of the zeros for the discrete cases manageable. As stated earlier, the bosonic
calculations had been done in Ref.~\cite{mulken01}, and we repeated some of
these to check our programs. Our calculations showed a clear tendency, 
even for as few as $N=10$ bosons, for Fisher zeros to
approach the real positive $\beta$ axis 
for $d=2$, and more so for $d=3$. This is shown in fig.~\ref{fig:0}
Although the 
actual distribution of zeros on the complex plane differed for the
discrete and the continuous density of states, both showed the
expected behaviour as $N$ was increased. These bosonic results
have been studied~\cite{mulken01}, and will not be displayed further.  
 
We now present the fermionic results. Consider first
the discrete density of states, which gives rise to $Z_1(\beta)$ of 
eq.~(\ref{sinh}), and the resulting $Z_N$ for fermions. 
\begin{figure}[htbp]
  \centering
\subfigure[~~Zeros of the partition function in two dimensions.]{
   \resizebox{3.0in}{!}{\includegraphics[angle=-90]{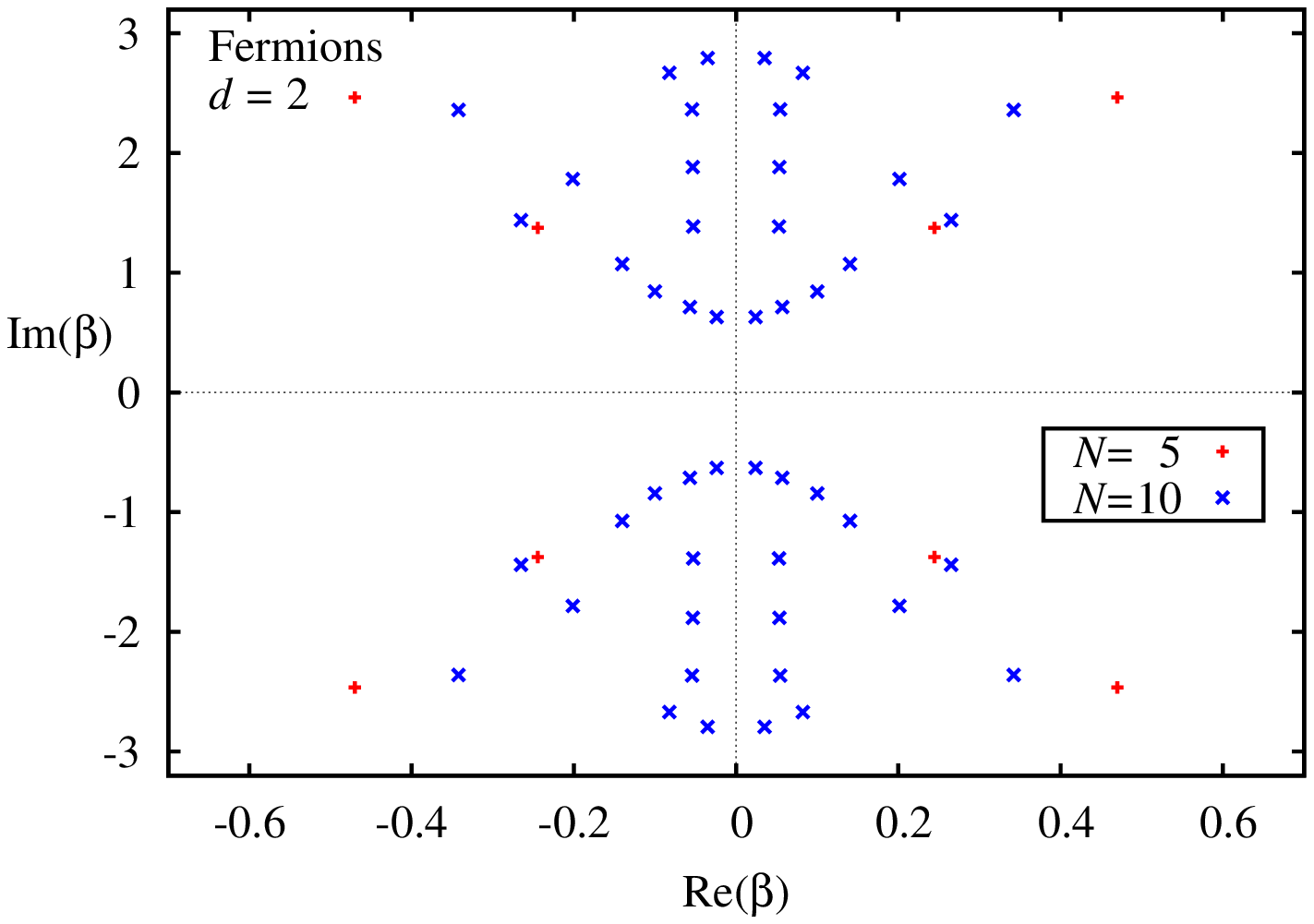}}
\label{subfig:1a}
\hspace{0.2in}
}
\subfigure[~~Zeros of the partition function in three dimensions.]{
   \resizebox{3.0in}{!}{\includegraphics[angle=-90]{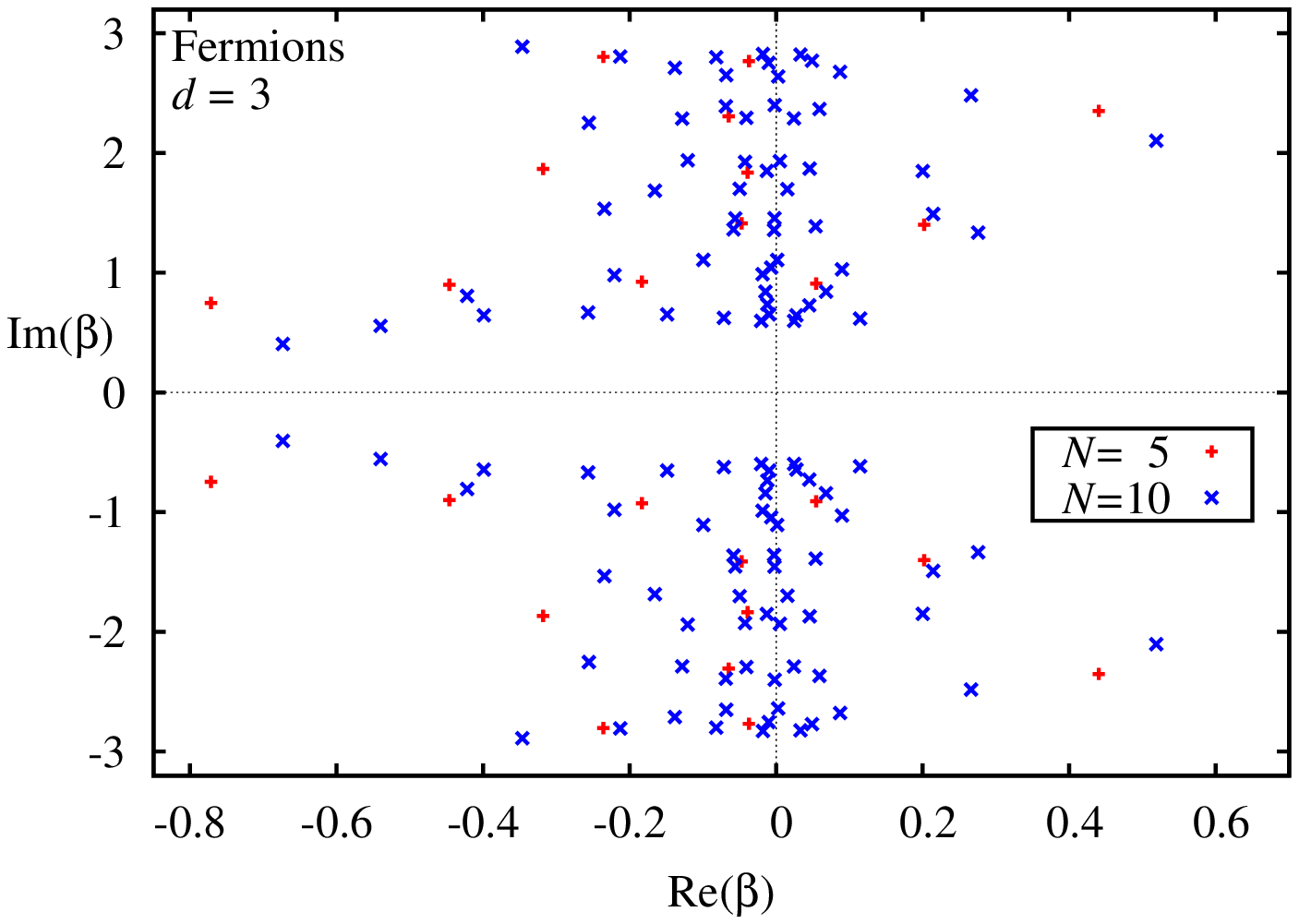}}
\label{subfig:1b}
}
 \caption{(Color online) The zeros of the exact $Z_N(\beta)$, for $N=5$ and $N=10$, obtained from $Z_1(\beta)$
  of eq.~(\ref{sinh}) as a function of $\beta=1/kT$ for noninteracting
  fermions in a harmonic oscillator well: 
(a) $d=2$  and (b) $d=3$.}
\label{fig:1}
\end{figure}
In fig.~\ref{fig:1}, 
its zeros on the complex plane are displayed for $N=5$ and $N=10$
fermions in two and three spatial dimensions.  There are no zeros other than the origin in one dimension.   Since $Z_N (\beta)$ is 
known, the thermal properties (for real $\beta$) are easily
calculated. These show the expected regular behaviour.  

\begin{figure}[htbp]
  \centering
\subfigure[~~Zeros of the partition function of 15 fermions for three different dimensions.]{
   \resizebox{3.0in}{!}{\includegraphics[angle=-90] 
{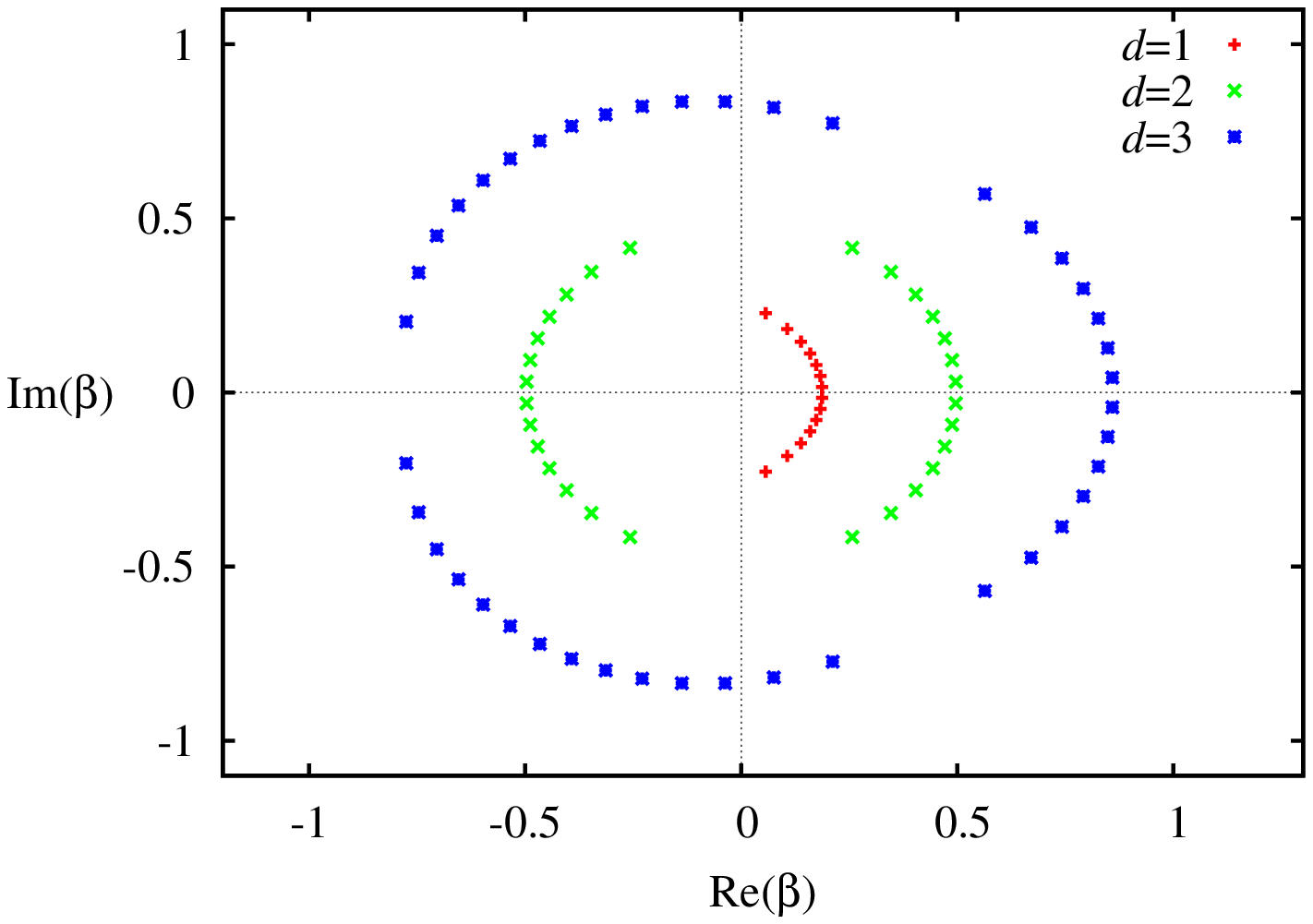}}
\label{subfig:3a}
\hspace{0.2in}
}
\subfigure[~~Heat capacity of a 15-fermion system.]{
   \resizebox{3.0in}{!}{\includegraphics[angle=-90]   
{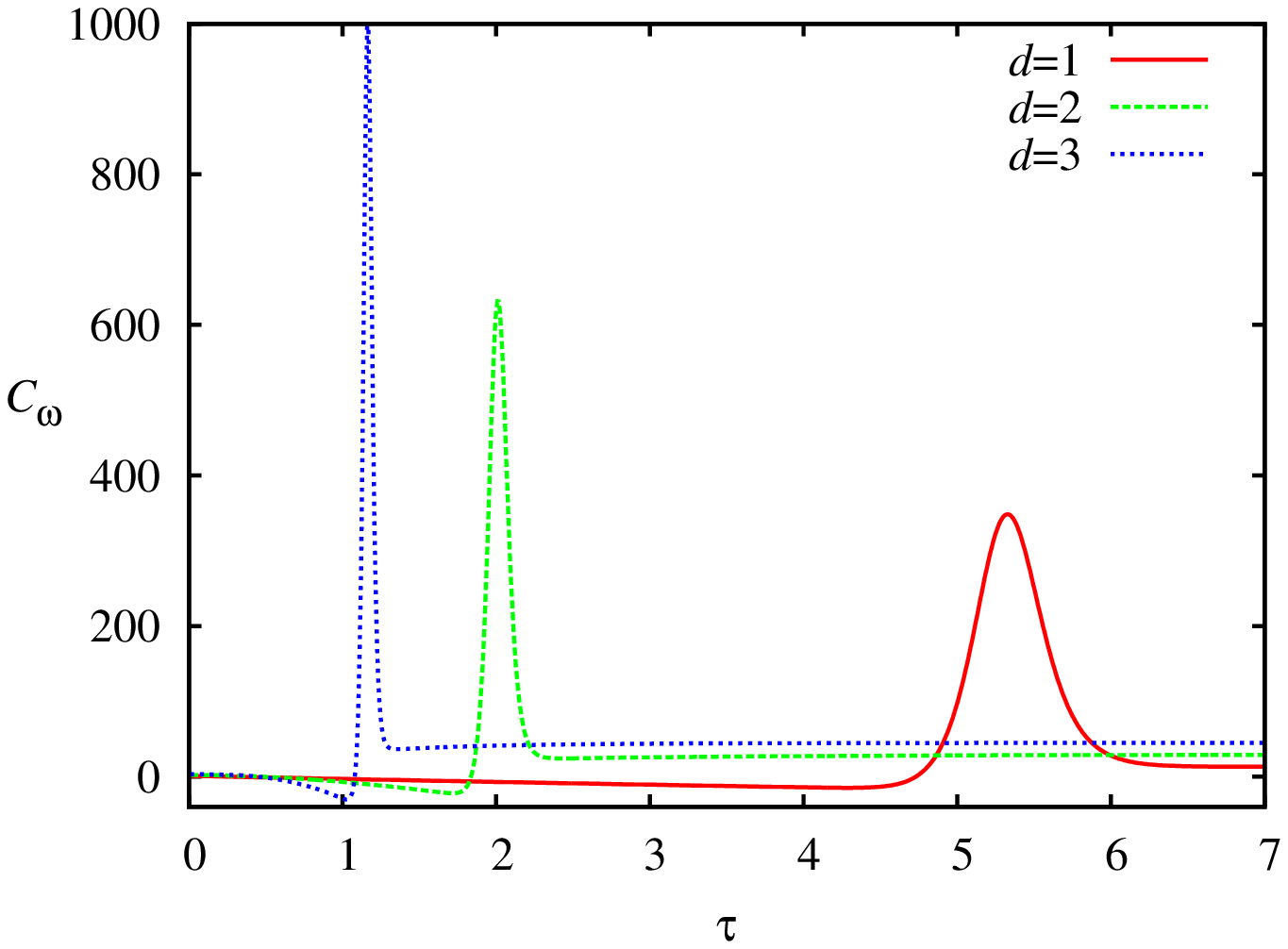}}
\label{subfig:3b}
}
\caption{(Color online) The zeros of the canonical partition function as a function of 
$\beta=1/k_BT$ (a) and the heat capacity as a function of $\tau=1/\beta$ (b) for a system of 15 noninteracting fermions with a continuous density of states, given by eq.~(\ref{smoothy}).}
\label{fig:3}
\end{figure}

When, however, $Z_N(\beta)$ is calculated using 
$\widetilde{Z}_1(\beta)$ given by eq.~(\ref{cont}) for the continuous 
density of states, we get very unexpected results. 
For even $N$, we find that one zero falls
precisely on the positive real $\beta$ axis, even for very small particle
number (i.e., $N=2$ ), regardless of the dimension $d$ . 
\begin{figure}[htbp]
  \centering
\subfigure[~~Zeros of $Z_N(\beta)$ for $10$ fermions calculated using 
  $\widetilde{Z}_1(\beta)$ given by eq.(\ref{cont}) in  
three different dimensions.]{
   \resizebox{3.0in}{!}{\includegraphics[angle=-90] 
{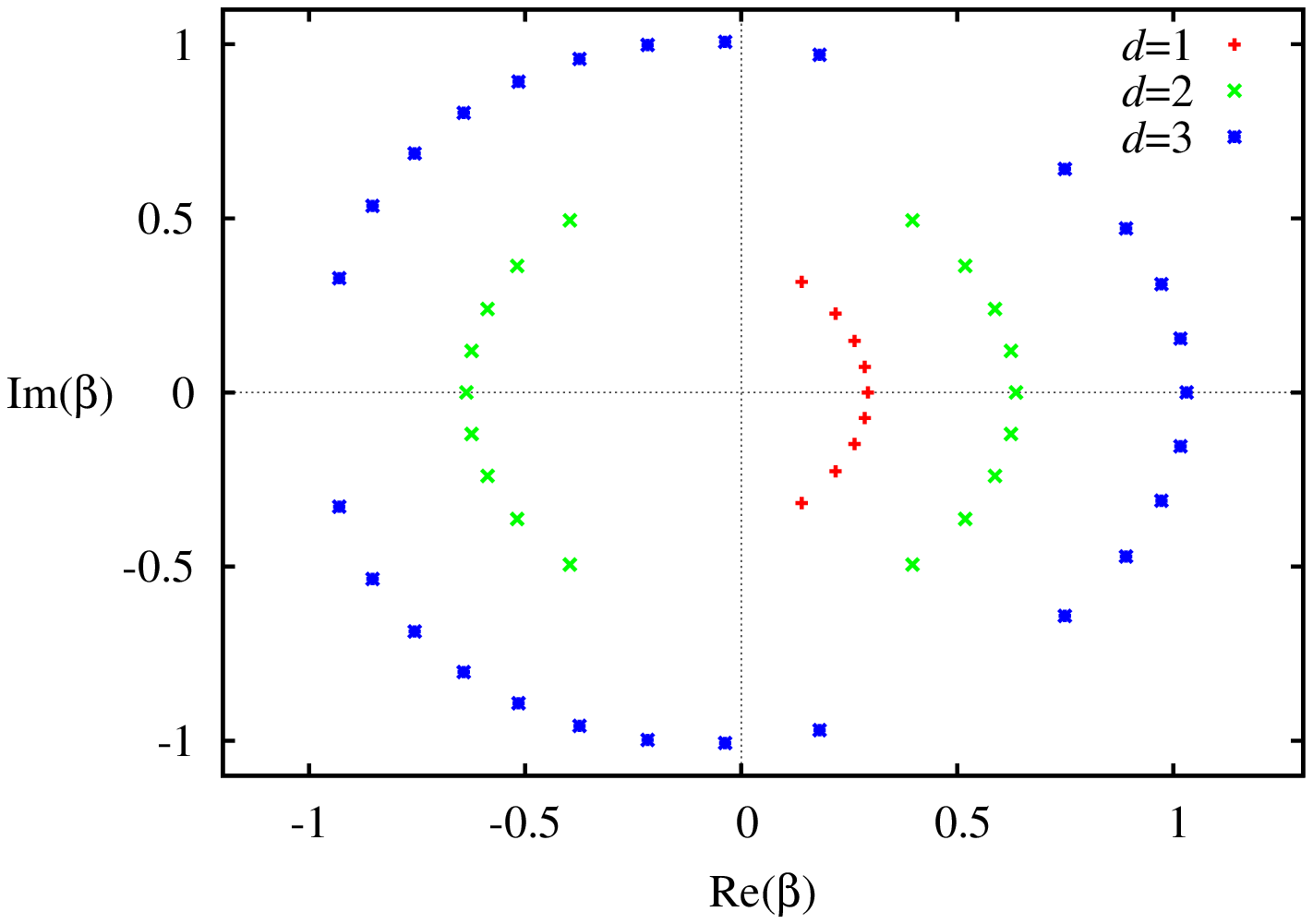}}
\label{subfig:2a}
\hspace{0.2in}
}
\subfigure[~~Heat capacity as a function of temperature for a
10-fermion 
system, under the same approximation as in (a) above.]{
   \resizebox{3.0in}{!}
   {\includegraphics[angle=-90]
{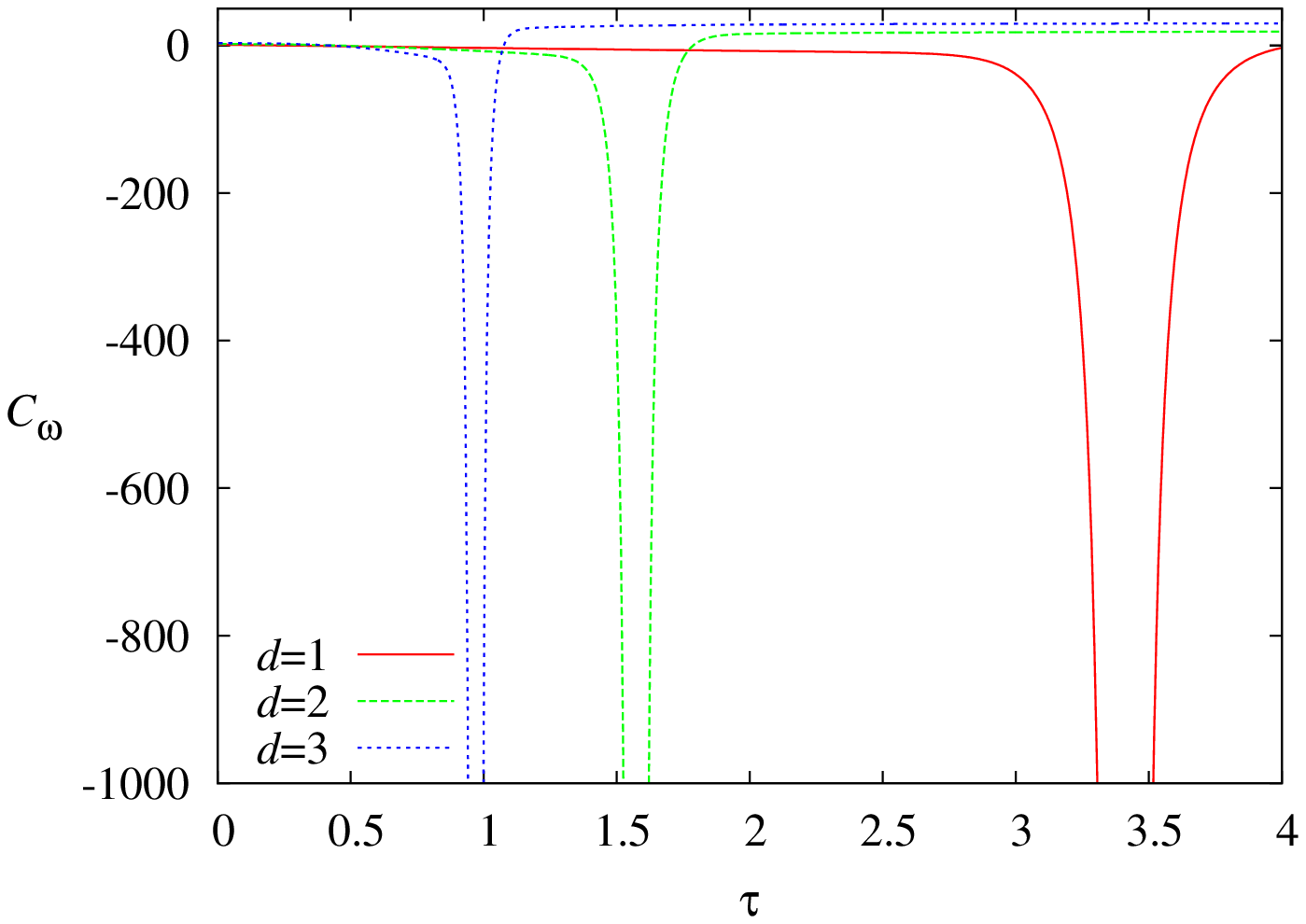}}
\label{subfig:2b}
}
\caption{(Color online) The zeros of the canonical partition function as a function of $\beta=1/kT$ (a) and the heat capacity as a function of $\tau=1/\beta$ (b) for a system of 10 noninteracting fermions with a continuous density of states, given by eq.~(\ref{smoothy}).}
\label{fig:2}
\end{figure}
We investigate the thermal properties of such a system when calculated 
using the canonical formalism. 
The resulting heat capacity is negative over a range of 
temperature, indicating instability of the system. This will be
discussed shortly. In fig.~\ref{fig:2}, 
we display the zeros of $Z_N(\beta)$ on the complex 
$\beta$ plane for $N=10$ in one, two and three spatial dimensions. 
The associated heat capacities for these cases are also shown. Similar 
behaviour is also found for very small to very large even $N$. 
For odd $N$, however, $Z_N(\beta)$ from the continuous density of states has no real zero on the positive $\beta$ axis.  This could be easily checked analytically, for example, for $N=3$.  Nevertheless, the complex zeros of $Z_N(\beta)$ get dense near the positive $\beta$ axis and give anomalous thermal properties.  This is displayed in fig.~\ref{fig:3} for $N=15$.  Similar results are obtained for smaller as well as larger odd $N$.
 
\begin{figure}[t]
  \centering
   \resizebox{3.2in}{!}{\includegraphics[angle=-90] 
{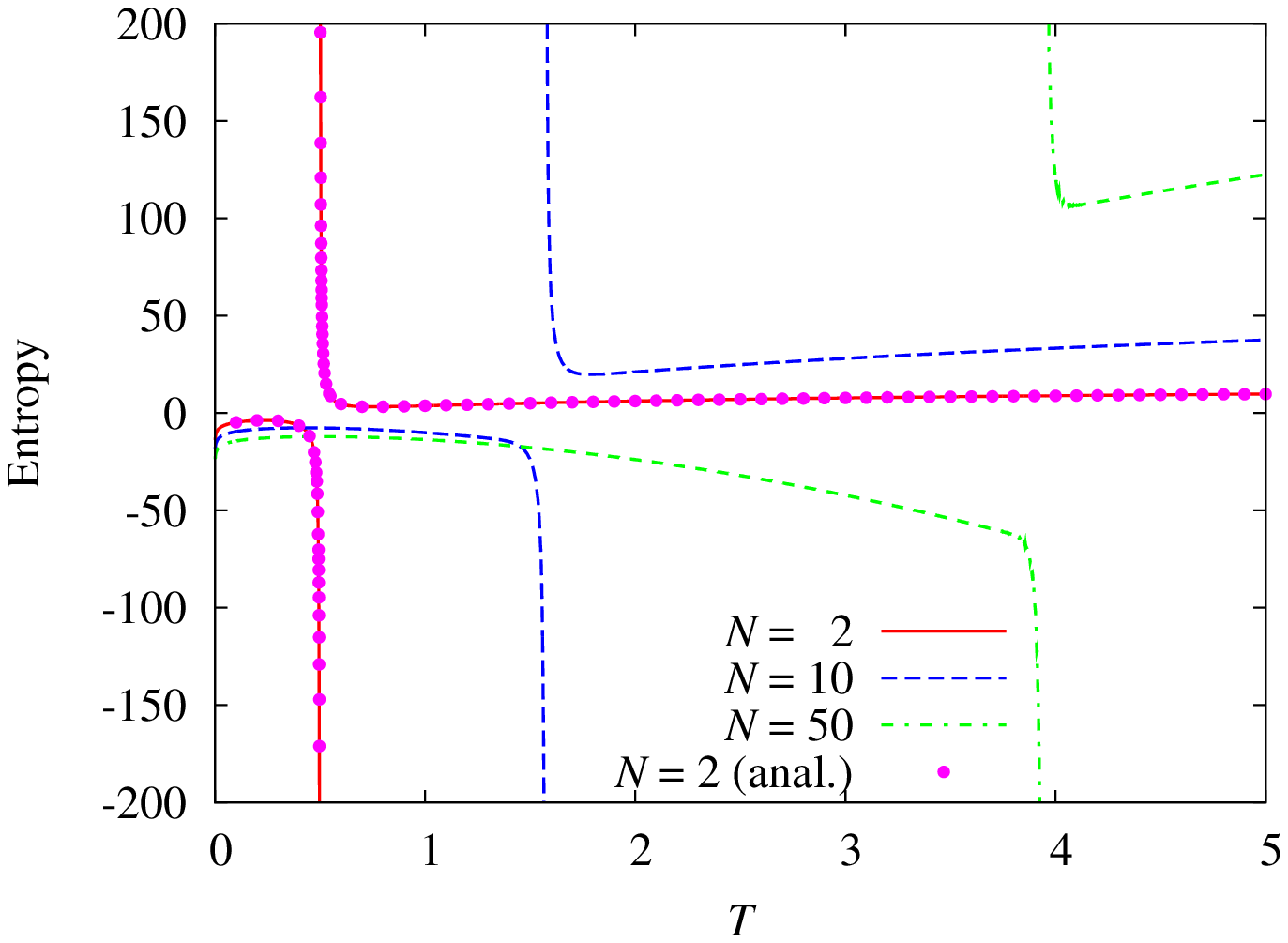}}
\caption{(Color online) Entropy $S$ for 2-dimensional fermion systems with continuous energy-density spectrum. $N=2$ (anal.) refers to entropy obtained from Eq.~(\ref{twice}).}
\label{fig:5}
\end{figure}

The present calculations with the continuous density of states give
unphysical results for fermions in a harmonic potential. 
For example, take the case of $N=2$. From Eq.~(\ref{party}), 
we get 
\beq
Z_2(\beta)=(Z_1^2(\beta)-Z_1(2\beta))/2~. 
\label{cook}
\eeq
For the continuous  
density of states, replacing $Z_1(\beta)$ by $\widetilde{Z}_1(\beta)$, and  
using Eq.~(\ref{cont}) for $d=2$ with $(\hbar\omega=1)$, we obtain 
\beq
\widetilde{Z}_2(\beta)=(1/\beta^4-1/(4\beta^2))/2~, 
\label{twice}
\eeq
which has a real 
positive root for $\beta=2$. This behaviour for trapped particles is 
spurious, and can be seen as follows. Consider the single-particle
canonical partition function of a one-dimensional system. The
$d$-dimensional counterpart may be obtained by taking its $d^\mathrm{th}$
 power. Writing $Z_1(\beta)=\sum_i\exp(-\beta \epsilon_i)$, we find, using 
Eq.(\ref{cook}) 
\begin{equation}\label{new}
\begin{split}
Z_2(\beta)&=\frac{1}{2}\left[\sum_i \sum_j e^{\textstyle -\beta (\epsilon_i+
\epsilon_j)}-\sum_i e^{\textstyle -2\beta\epsilon_i}\right] 
\\
          &=\sum_i\sum_{j<i}e^{\textstyle -\beta(\epsilon_i+\epsilon_j)}~.
\end{split}          
\end{equation}
The second term on the RHS of Eq.~(\ref{cook}) subtracts out the
contributions where the two particles occupy the same orbital. 
The resulting expression for $Z_2(\beta)$ in Eq.~(\ref{new}) above is 
in accordance with the Pauli principle, and 
cannot be zero for real positive $\beta\neq 0 $, if all states have 
discrete eigenvalues. 
The roots of $\widetilde{Z}_2(\beta)=0$ found from Eq.~(\ref{twice}) may not be
physically meaningful. This analysis may be extended for larger (finite) $N$.
Figure~\ref{fig:5} shows the variation of entropy $S$ with temperature for
$N=2$, $N=10$ and $N=50$ fermions trapped in a  $d=2$ harmonic
oscillator, using the continuous
density of states given by Eq.~(\ref{smoothy}). The behaviour in all
three cases is the same, and is dominated by an unphysical pole in $S$  
due to the anomalous Fisher zero.  Note that the anomalous zero shifts from $T=1/2$ to about $T=4$ as $N$ varies from 2 to 50.   For $N=2$, the anomalous zero is at 
$T=1/2$, and near this $T$, $S\simeq (4T^2-1/2)/(T^2-1/4)$. This
causes $S(T)$ to plunge to large negative values for $T<1/2$, giving
rise to a discontinuity at $T=1/2$. This explains why the heat
capacity is negative in this range of $T$. 

The energy fluctuation in a 
canonical ensemble is $(\langle E^2\rangle - \langle E\rangle^2) =T^2 C_{\omega}(T)$, where 
$\langle\dots\rangle$ denotes an ensemble average. Since the energy fluctuation may 
also be written as $\langle(E- \langle E \rangle)^2\rangle$, we normally expect this to be
positive. The change in sign can, however, be explained by examining 
the inverse Laplace transform of Eq.~(\ref{twice}), which gives the 
two-fermion continuous density of states. This is not positive
definite, and ensemble averaging $\langle\dots\rangle$ is no longer guaranteed to
yield a positive definite answer. This is easily checked, for example,     
by calculating, for $E>0$, the ensemble average $\langle E\rangle$ as a function of 
$T$.

Quite generally, the exact density of states for particles moving in a 
potential, consists of a smooth continuous part, and a sum of
oscillating terms~\cite{brack03}. Equation~(\ref{osc}) for the HO is a
specially simple example of this. It is the oscillatory contribution  
that gives rise to the discrete delta-function spikes to the density
of states. 
The oscillatory part has a large effect
on the analytical behaviour of $Z_N(\beta)$, specially for fermions. 
The fermionic $Z_N(\beta)$ is more sensitive since the recurrence
relation given by Eq.~(\ref{party}) has alternating positive and
negative terms. This is unlike the situation in the grand canonical partition
function, given by 
\begin{equation}
\ln Z_G(\alpha,\beta)=\int_0^{\infty} g(\epsilon)\ln\big(1+\exp(\alpha-\beta \epsilon)\big)d\epsilon~, 
\end{equation}
where 
$\alpha=\beta \mu $, and $\mu$ the chemical potential. It can be
shown that the contribution of the oscillatory terms is exponentially
small with a $1/\beta$ dependence in the exponent~\cite{richter96}. 
The grand canonical formalism generally gives sensible results using
the smooth single-particle density of states $\widetilde{g}(\epsilon)$. 
We show that the same $\widetilde{g}(\epsilon)$, when used recursively to 
generate the canonical $\widetilde{Z}_N(\beta)$, yields anomalous results 
subject to misinterpretation. 

\FloatBarrier
\acknowledgements
R.K.B. acknowledges useful discussions with Jules Carbotte.  W.v.D. acknowledges financial support from the Natural Sciences and Engineering Research Council of Canada. 


\begin{thebibliography}{10}
\expandafter\ifx\csname url\endcsname\relax\def\url#1{\texttt{#1}}\fi

\bibitem{yang52}
\Name{Yang C.~N. \and Lee T.~D.} \REVIEW{Phys. Rev.}{87}{1952}{404}.

\bibitem{lee52}
\Name{Lee T.~D. \and Yang C.~N.} \REVIEW{Phys. Rev.}{87}{1952}{410}.

\bibitem{fisher65}
\Name{Fisher M.~E.} in \Book{Lectures in theoretical physics}, edited by
  \Name{Britten W.} Vol.~7C (University of Colorado Press, Boulder, Colorado)
  1965 pp. 42--45.

\bibitem{borrmann93}
\Name{Borrmann P. \and Franke G.} \REVIEW{J. Chem. Phys.}{98}{1993}{2484}.

\bibitem{mulken01}
\Name{M\"ulken O., Borrmann P., Harting J. \and Stamerjohanns H.} \REVIEW{Phys.
  Rev. A}{64}{2001}{013611}.

\bibitem{janke01}
\Name{Janke W. \and Kenna R.} \REVIEW{J. Stat. Phys.}{102}{2001}{1211}.

\bibitem{janke02}
\Name{Janke W. \and Kenna R.} \REVIEW{Nucl. Phys. B (Proc.
  Suppl.)}{106-107}{2002}{905}.

\bibitem{das05}
\Name{Das C., Das Gupta S., Lynch W., Mekjian A. \and Tsang M.} \REVIEW{Phys.
  Rep.}{406}{2005}{1}.

\bibitem{kenna00}
\Name{Kenna R., Pinto C. \and Sexton J.} \REVIEW{Nucl. Phys. B (Proc.
  Suppl.)}{83}{2000}{667}.

\bibitem{baille87}
\Name{Baille C.} \REVIEW{Nucl. Phys. B}{283}{1987}{217}.

\bibitem{brack03}
\Name{Brack M. \and Bhaduri R.~K.} \Book{Semiclassical physics} (Westview
  Press, Colorado) 2003 p. 122.

\bibitem{philips57}
\Name{Phillips E.} \Book{Functions of a complex variable} 8th Edition (Oliver
  and Boyd, London) 1957.

\bibitem{abramowitz65}
\Name{Abramowitz M. \and Stegun I.~A.} \Book{Handbook of mathematical
  functions} (Dover Publications, Inc., New York) 1965 p. 85.

\bibitem{schmidt99}
\Name{Schmidt H.~J. \and Schnack J.} \REVIEW{Physica A}{265}{1999}{584}.

\bibitem{richter96}
\Name{Richter K., Ullmo D. \and Jalabert R.~A.} \REVIEW{Phys.
  Rep.}{276}{1996}{1}.

\end{thebibliography}

\end{document}